\documentclass[11pt]{article}
\usepackage{amsmath,amssymb,amsthm,amsxtra}
\usepackage{overpic}
\begin{document}

\vspace{0.2cm}

\begin{center}
{\Large\bf The Kobayashi-Maskawa Parametrization of Lepton Flavor Mixing
and Its Application to Neutrino Oscillations in Matter}
\end{center}

\vspace{0.1cm}
\begin{center}
{\bf Ye-Ling Zhou}
\footnote{E-mail: zhouyeling@ihep.ac.cn} \\
{\sl Institute of High Energy Physics,
Chinese Academy of Sciences, Beijing 100049, China}
\end{center}

\vspace{1.5cm}

\begin{abstract}
We show that the Kobayashi-Maskawa (KM) parametrization of the
$3\times 3$ lepton flavor mixing matrix is a useful language to
describe the phenomenology of neutrino oscillations. In particular,
it provides us with a convenient way to link the genuine flavor
mixing parameters ($\theta^{}_1$, $\theta^{}_2$, $\theta^{}_3$ and
$\delta^{}_{\text{KM}}$) to their effective counterparts in matter
($\tilde{\theta}^{}_1$, $\tilde{\theta}^{}_2$, $\tilde{\theta}^{}_3$
and $\tilde{\delta}^{}_{\text{KM}}$). We rediscover the Toshev-like
relation $\sin\tilde{\delta}_{\text{KM}}\sin2\tilde{\theta}^{}_2=
\sin\delta^{}_{\text{KM}}\sin2\theta^{}_2$ in the KM
parametrization. We make reasonable analytical approximations to the
exact relations between the genuine and matter-corrected flavor
mixing parameters in two different experimental scenarios: (a) the
neutrino beam energy $E$ is above ${\cal O}(1)$ GeV and (b) $E$ is
below ${\cal O}(1)$ GeV. As an example, the probability of
$\nu^{}_\mu \rightarrow \nu^{}_e$ oscillations and CP-violating
effects are calculated for the upcoming NO$\nu$A and Hyper-K
experiments.
\end{abstract}

\begin{flushleft}
\hspace{0.8cm} PACS number(s): 14.60.Pq, 13.10.+q, 25.30.Pt
\end{flushleft}

\newpage

\section{Introduction}
Just like quark flavor mixing, lepton flavor mixing has been
observed in a number of neutrino oscillation experiments \cite{PDG}.
The phenomenon of lepton flavor mixing is described by a $3\times 3$
unitary matrix $V$, the so-called Maki-Nakawaga-Sakata-Pontecorvo
(MNSP) matrix \cite{MNSP}. There are totally 9 different
parametrizations of $V$ in terms of the rotation angles and phase
angles \cite{FXzz}. Among them, the one advocated by the Particle
Data Group (PDG) \cite{PDG} is most popular in accounting for
current neutrino oscillation data, and the Fritzsch-Xing (FX)
parametrization \cite{FXzz} has a particular merit in describing the
running behaviors of neutrino masses and flavor mixing parameters
from one energy scale to another by means of the one-loop
renormalization-group equations \cite{Xing2006}. Is the original
Kobayashi-Maskawa (KM) parametrization \cite{KM} advantageous to the
description of neutrino phenomenology? We shall give an affirmative
answer to this question in the present work.

In fact, it has recently been noticed that the KM parametrization of
the $3\times3$ quark flavor mixing matrix is very useful to link
current experimental data to the unitarity triangles--- the
so-called ``Unitarity boomerang" \cite{He}: $\delta^{}_{\rm KM}
\simeq \alpha^{}_{\rm UT} \simeq 90^\circ$, where $\delta^{}_{\rm
KM}$ is the CP-violating phase in the KM parametrization and
$\alpha^{}_{\rm UT}$ is one of the inner angles of the KM unitarity
triangles. A similar relationship was found earlier in the FX
parametrization \cite{XingZz}. As the structure of the KM
parametrization is partly analogous to that of the FX
parametrization and partly analogous to that of the PDG
parametrization, we naturally expect that it should also be useful
to describe the salient features of neutrino oscillations in vacuum
and in matter. In other words, we expect that the KM parametrization
can provide us with a simple and convenient link between its
parameters and the observable quantities of neutrino oscillations.
The main purpose of this paper is just to demonstrate our
expectation and offer an alternative description of lepton flavor
mixing for both the phenomenology of neutrino oscillations and the
building of neutrino mass models. Needless to say, a convenient
parametrization is sometimes possible to make the underlying physics
more transparent.

The remaining parts of this work are organized as follows:

(1) In section 2 we shall first establish the explicit relations
between the flavor mixing parameters in the KM parametrization
($\theta^{}_1$, $\theta^{}_2$, $\theta^{}_3$ and
$\delta^{}_\text{KM}$) and those in the PDG parametrization
($\theta^{}_{12}$, $\theta^{}_{13}$, $\theta^{}_{23}$ and $\delta$).
Because the Majorana CP-violating phases have nothing to do with
neutrino oscillations, they will not be taken into account in this
work. We find that $\theta^{}_2 = \theta^{}_{23}$ and
$\delta^{}_\text{KM}=\delta$ hold exactly if the condition
$\cos\delta=\sin\theta^{}_{13} \cot\theta^{}_{12}
\cot2\theta^{}_{23}$ (or equivalently
$\cos\delta^{}_\text{KM}=\tan\theta^{}_3 \cos\theta^{}_1
\cot2\theta^{}_2$) is satisfied. Because both $\theta^{}_{13}$ and
$\theta^{}_3$ are expected to be small, the above condition seems to
hint at $\delta \sim \delta^{}_{\rm KM} \sim \pm 90^\circ$.

(2) Section 3 is devoted to a detailed calculation of the relations
between the genuine KM flavor mixing parameters in vacuum
($\theta^{}_1$, $\theta^{}_2$, $\theta^{}_3$ and
$\delta^{}_\text{KM}$) and their effective counterparts in matter
($\tilde{\theta}^{}_1$, $\tilde{\theta}^{}_2$, $\tilde{\theta}^{}_3$
and $\tilde{\delta}^{}_\text{KM}$), given neutrino oscillations in a
constant terrestrial matter profile. The so-called Toshev relation
$\sin\tilde{\delta}\sin2\tilde{\theta}^{}_{23}
=\sin\delta\sin2\theta^{}_{23}$ \cite{Toshev} in the PDG
parametrization is rediscovered in the KM parametrization:
$\sin\tilde{\delta}_\text{KM}\sin2\tilde{\theta}^{}_2=
\sin\delta^{}_\text{KM}\sin2\theta^{}_2$. This interesting result
means that the KM parametrization is definitely useful and
convenient to describe the phenomenology of neutrino oscillations.
To be more explicit, we make analytical approximations for the
relations between the genuine and matter-corrected parameters of
lepton flavor mixing in two different experimental scenarios: (a)
the neutrino beam energy $E$ is above ${\cal O}(1)$ GeV; and (b) $E$
is below ${\cal O}(1)$ GeV. The accuracy of each approximation is
examined by comparing its results with exact numerical calculations.
Such analytical results are phenomenologically useful, just like
those obtained previously in the PDG parametrization.

(3) For illustration, we consider $\nu^{}_\mu \to \nu^{}_e$
oscillations in section 4 and calculate the oscillation probability
by means of the KM parametrization. Our expressions are simple and
instructive, and they can be used to analyze the upcoming data from
T2K \cite{T2K}, NO$\nu$A \cite{NOvA} and Hyper-K \cite{HyperK}
experiments.

(4) A brief summary of this work, together with some concluding
remarks, is given in section 5.

\section{Comparison between the KM and PDG parametrizations}

In the framework of 3-generation leptons, the KM and PDG
parametrizations of the MNSP matrix are given by
\begin{eqnarray}
&&V^{}_{(\text{KM})}=\begin{pmatrix}c^{}_1 &-s^{}_1c^{}_3
&-s^{}_1s^{}_3
\\s^{}_1c^{}_2 &c^{}_1c^{}_2c^{}_3-s^{}_2s^{}_3e^{i\delta^{}_\text{KM}} &
c^{}_1c^{}_2s^{}_3+s^{}_2c^{}_3e^{i\delta^{}_\text{KM}}
\\s^{}_1s^{}_2& c^{}_1s^{}_2c^{}_3+c^{}_2s^{}_3e^{i\delta^{}_\text{KM}}
& c^{}_1s^{}_2s^{}_3-c^{}_2c^{}_3e^{i\delta^{}_\text{KM}}
\end{pmatrix}\;,\nonumber\\
&&V^{}_{(\text{PDG})}=\begin{pmatrix} c^{}_{12}c^{}_{13}&s^{}_{12}c^{}_{13}&s^{}_{13}e^{-i\delta}\\
-s^{}_{12}c^{}_{23}-c^{}_{12}s^{}_{23}s^{}_{13}e^{i\delta}&s^{}_{12}c^{}_{23}
-s^{}_{12}s^{}_{23}s^{}_{13}e^{i\delta} &s^{}_{23}c^{}_{13}\\
s^{}_{12}s^{}_{23}-c^{}_{12}c^{}_{23}s^{}_{13}e^{i\delta}&
-c^{}_{12}s^{}_{23}-s^{}_{12}c^{}_{23}s^{}_{13}e^{i\delta}&c^{}_{23}c^{}_{13}
\end{pmatrix}\;,
\end{eqnarray}
respectively, where $s^{}_i=\sin\theta^{}_i$,
$c^{}_i=\cos\theta^{}_i$, $s^{}_{ij}=\sin\theta^{}_{ij}$ and
$c^{}_{ij}=\cos\theta^{}_{ij}$ (for $i=1,2,3$ and $ij=12,23,13$).
Comparing the KM parametrization with the PDG parametrization, one
can derive the expressions of $\theta^{}_i$ and
$\delta^{}_\text{KM}$ in terms of $\theta^{}_{ij}$ and $\delta$:
\begin{eqnarray}\label{kmpdg}
&&\cos  \theta^{}_1=\cos  \theta^{}_{12} \cos  \theta^{}_{13}\;,\nonumber\\
&&\tan  \theta^{}_2=\tan  \theta^{}_{23}\left|\frac{1- \sin  \theta
_{13}\cot
 \theta^{}_{12} \cot  \theta^{}_{23} e^{\text{i$\delta $}}}{1+
  \sin  \theta^{}_{13}\cot  \theta^{}_{12}\tan  \theta^{}_{23}
  e^{\text{i$\delta $}}}
  \right|\;,\nonumber\\
&&\tan  \theta^{}_3=\tan  \theta^{}_{13}\csc  \theta^{}_{12}\;,\nonumber\\
&&\sin  \delta^{}_{\text{KM}}=\sin  \delta  \frac{1+ \sin ^2 \theta
_{13}\cot ^2 \theta^{}_{12} }{\left|\left(1-\sin  \theta^{}_{13}
\cot \theta^{}_{12} \cot  \theta^{}_{23} e^{\text{i$\delta
$}}\right) \left( 1+ \sin  \theta^{}_{13}\cot  \theta^{}_{12}\text{
}\tan  \theta^{}_{23}e^{\text{i$\delta $}}\right)\right| }\;.
\end{eqnarray}
With the help of Eq. \eqref{kmpdg} and the current experimental
oscillation data \cite{Schwetz}, we find the following properties of
$\theta^{}_i$ and $\delta^{}_\text{KM}$:
\begin{itemize}
\item $\theta^{}_{1}\approx \theta^{}_{12}$ is a good approximation
because of $\theta^{}_{13}\lesssim 12^\circ_{}$ as constrained by
the present experimental data \cite{Schwetz}.

\item $\theta^{}_2$ depends on not only  $\theta^{}_{23}$ but also
$\theta^{}_{13}$ and $\delta$. $\theta^{}_2\approx\theta^{}_{23}$ is
a good approximation if $\delta$ is near $\pm90^\circ_{}$, but not
good if $\delta$ is near $0^\circ_{}$ or  $\pm180^\circ_{}$. If one
requires $\theta^{}_2 =\theta^{}_{23}$ to hold exactly and
$\theta^{}_{13}\neq0$, a constraint equation in the PDG
parametrization must be satisfied:
\begin{eqnarray}\label{ceq}
\cos\delta=\sin\theta^{}_{13}\cot\theta^{}_{12}
\cot2\theta^{}_{23}\;.
\end{eqnarray}

\item  When calculating the effective
flavor mixing parameters and neutrino oscillation probabilities in
matter, one of ten takes $\theta^{}_{13}$ as an expansion parameter
to make analytical approximations because of its small value ({\it
e.g.}, \cite{CFA}). Since $\tan\theta^{}_3 =
\tan\theta^{}_{13}\csc\theta^{}_{12}$ holds, we can also treat
$\theta^{}_3$ as a small parameter to do series expansions.

\item The experimental data yield $\theta^{}_{23}\approx 45^\circ_{}$ \cite{Schwetz},
leading to $\displaystyle\sin \delta^{}_{\text{KM}}\approx \sin
\delta+\mathcal{O}(s^2_{13})$ from Eq. \eqref{kmpdg}. So
$\delta^{}_\text{KM}\approx\delta$ is a good approximation. Under
the condition of Eq. \eqref{ceq}, $\delta^{}_\text{KM}=\delta$ holds
exactly.
\end{itemize}
We remark that Eq. \eqref{ceq} is a useful constraint that leads
exactly to both $\theta^{}_2 = \theta^{}_{23}$ and
$\delta^{}_\text{KM}=\delta$. It can be re-expressed in the KM
parametrization:
\begin{eqnarray}\label{ceqKM}
\cos\delta^{}_\text{KM}=\tan\theta^{}_3 \cos\theta^{}_1
\cot2\theta^{}_2 \;.
\end{eqnarray}
Moreover, one can see from Eqs. \eqref{ceq} and \eqref{ceqKM} that
$\theta^{}_2=\theta^{}_{23}$ is equivalent to
$\delta^{}_\text{KM}=\delta$. One special case is the $\mu$-$\tau$
flavor symmetry with $\theta^{}_2=\theta^{}_{23}=45^{\circ}_{}$ and
maximal CP violation with
$\delta^{}_\text{KM}=\delta=\pm90^{\circ}_{}$ \cite{mutau}. We shall
consider this possibility in the next section when discussing matter
effects.

We obtain the best-fit values and the $1\sigma$, $2\sigma$ and
$3\sigma$ ranges of mixing angles $\theta^{}_1$, $\theta^{}_2$ and
$\theta^{}_3$ for both the normal hierarchy (NH) and inverted
hierarchy (IH) of neutrino masses in Table 1, according to a global
analysis of current neutrino data presented in Ref. \cite{Schwetz}.
Some details of this table should be mentioned. First, we do not
list any possible values of the CP phase $\delta^{}_\text{KM}$
because $\delta^{}_\text{KM}$ is not well constrained by current
experiments. Ref. \cite{Schwetz} gives very loose bounds of
$\delta$, with $\delta=-110^\circ_{}$ ($-74^\circ_{}$) for the NH
(IH) at the best fit and $-227^\circ_{}$ to $+7^\circ_{}$
($-200^\circ_{}$ to $+43^\circ_{}$) for the NH (IH) in the $1\sigma$
range, without constraints in the $2\sigma$ or $3\sigma$ ranges. In
this case, one may assume the bounds of $\delta$ as the bounds of
$\delta^{}_\text{KM}$ for $\delta^{}_\text{KM}\approx\delta$.
Second, because we have little knowledge about $\delta$,
$\theta^{}_2$ cannot be well restricted from Eq. \eqref{kmpdg}. Thus
we assume $\delta$ at its best-fit value when we calculate the range
of $\theta^{}_2$ in Table 1. For a full understanding of the
possible values of $\theta^{}_2$ and their dependence on $\delta$,
one may see Fig. 1, where we take the NH as an example to show how
$\theta^{}_2$ changes with $\delta$.

\section{The KM flavor mixing parameters in matter}

\subsection{General formalism}

To understand the phenomenon of neutrino oscillations in the
long-baseline experiments, it is necessary to analyze neutrino
mixing in matter. In this section, we calculate the effective flavor
mixing parameters in a constant terrestrial matter profile
($\tilde{\theta}^{}_1$, $\tilde{\theta}^{}_2$, $\tilde{\theta}^{}_3$
and $\tilde{\delta}^{}_\text{KM}$) and study their relations with
the genuine flavor mixing parameters in vacuum ($\theta^{}_1$,
$\theta^{}_2$, $\theta^{}_3$ and $\delta^{}_\text{KM}$).

In the flavor basis $|\nu(t)\rangle \equiv \left(
|\nu^{}_e(t)\rangle\;, |\nu^{}_\mu(t)\rangle\;,
|\nu^{}_\tau(t)\rangle \right)^T_{}$, the evolution of neutrinos in
matter is described by a Schr$\ddot{\text{o}}$dinger-like equation:
\begin{eqnarray}
i\frac{\text{d}}{\text{d}t} |\nu(t)\rangle = \tilde{H}
|\nu(t)\rangle
\end{eqnarray}
with the effective Hamiltonian
\begin{eqnarray}\label{H}
\tilde{H}=\frac{\Delta m^2_{31}}{2E}\left[V\begin{pmatrix}0&0&0\\0&\alpha&0\\
0&0&1\end{pmatrix}V^\dag
+\begin{pmatrix}A&0&0\\0&0&0\\0&0&0\end{pmatrix}\right].
\end{eqnarray}
Here $V$ is the MNSP matrix, $\alpha\equiv\Delta m^2_{21}/\Delta
m^2_{31}$ is the mass hierarchy parameter with the mass-squared
differences $\Delta m^2_{ij}\equiv m^2_i-m^2_j$, and $A\equiv 2 E
a/\Delta m^2_{31}$ is a dimensionless variable arising from the
matter-induced effective potential $a\equiv\sqrt{2} G^{}_F N^{}_e$
\cite{matter}. $N^{}_e$ is the number density of electrons in
matter, and it can be taken to be half of the number density of
nucleons of the Earth. In this case, $A$ is given by
\begin{eqnarray}\label{A}
A \approx 0.085 \left(\frac{2.5\times
10^{-3}_{}\;\text{eV}^2}{\Delta m^2_{31}}\right)
\left(\frac{E}{1\;\text{GeV}}\right)
\left(\frac{\rho^{}_\text{m}}{2.8\;\text{g}/\text{cm}^3}\right)\;,
\end{eqnarray}
where $\rho^{}_\text{m}$ is the mass density along the path of
neutrinos. In most long-baseline neutrino oscillation experiments,
$\rho^{}_\text{m}$ is approximately a constant
\cite{T2K,NOvA,HyperK}. Eq. \eqref{H} holds for neutrinos. When
considering the evolution of antineutrinos, we have to perform the
replacements $V\Rightarrow V^*_{}$ and $a\Rightarrow -a$ in the
effective Hamiltonian.

In view of Eq. \eqref{H}, one may define the effective MNSP matrix
$\tilde{V}$ through
\begin{eqnarray}
\tilde{H}\equiv\frac{\Delta m^2_{31}}{2E}\tilde{V}\begin{pmatrix}\lambda^{}_1&0&0 \\0&\lambda^{}_2&0\\
0&0&\lambda^{}_3\end{pmatrix}\tilde{V}^\dag \;,
\end{eqnarray}
where $\lambda^{}_i$ are the eigenvalues of the matrix in the square
bracket of Eq. \eqref{H}. The effective matter-corrected
mass-squared differences are written as $\Delta
\tilde{m}^2_{21}=\Delta m^2_{31}
\left(\lambda^{}_2-\lambda^{}_1\right)$ and $\Delta
\tilde{m}^2_{31}=\Delta m^2_{31}
\left(\lambda^{}_3-\lambda^{}_1\right)$.

To describe the MNSP matrices $V$ and $\tilde{V}$ in the KM
parametrization, we need three real rotation matrices and one
diagonal phase matrix $O^{}_1$, $O^{}_2$, $O^{}_3$ and
$U^{}_\delta$:
\begin{eqnarray}
O^{}_1=\begin{pmatrix}c^{}_1&-s^{}_1&0\\s^{}_1&c^{}_1&0\\0&0&1\end{pmatrix}\;,\;\;
O^{}_2=\begin{pmatrix}1&0&0\\0&c^{}_2&-s^{}_2\\0&s^{}_2&c^{}_2\end{pmatrix}\;,\;\;
O^{}_3=\begin{pmatrix}1&0&0\\0&c^{}_3&s^{}_3\\0&-s^{}_3&c^{}_3\end{pmatrix}\;,\;\;
U^{}_\delta=\begin{pmatrix}1&0&0\\0&1&0\\0&0&-e^{i\delta^{}_\text{KM}}_{}\end{pmatrix}\;,\nonumber
\end{eqnarray}
in which $O^{}_1$ is a rotation matrix in the (1,2) plane and
$O^{}_2$ and $O^{}_3$ are rotation matrices in the (2,3) plane. Thus
$V$ and $\tilde{V}$ can be parametrized as
\begin{eqnarray}
&&V=O^{}_2U^{}_\delta O^{}_1O^{}_3\;,\nonumber\\
&&\tilde{V}=\tilde{O}^{}_2\tilde{U}^{}_\delta\tilde{O}^{}_1\tilde{O}^{}_3\;,\label{Vt}
\end{eqnarray}
with the flavor mixing parameters in vacuum ($\theta^{}_1$,
$\theta^{}_2$, $\theta^{}_3$ and $\delta^{}_\text{KM}$)  and in
matter ($\tilde{\theta}^{}_1$, $\tilde{\theta}^{}_2$,
$\tilde{\theta}^{}_3$ and $\tilde{\delta}^{}_\text{KM}$),
respectively.

In the representation of the rotation matrices and phase matrix, Eq.
\eqref{H} can be rewritten as
\begin{eqnarray}
\tilde{H}=\frac{\Delta m^2_{31}}{2E}O^{}_2 U^{}_\delta M
U^\dag_\delta O^T_2
\end{eqnarray}
with
\begin{eqnarray}\label{H1}
M=O^{}_1 O^{}_3
\begin{pmatrix}0&0&0\\0&\alpha&0\\0&0&1\end{pmatrix}O^T_3
O^T_1 +\begin{pmatrix}A&0&0\\0&0&0\\0&0&0\end{pmatrix}\;.
\end{eqnarray}
Being a real symmetric matrix, $M$ can be diagonalized by an
orthogonal matrix $\hat{V}\equiv
\hat{O}^{}_2\hat{O}^{}_1\hat{O}^{}_3$ with three rotation angles
$\hat{\theta}^{}_1$, $\hat{\theta}^{}_2$ and $\hat{\theta}^{}_3$.
Thus, $\tilde{H}$ is diagonalized by $\tilde{V}'\equiv O^{}_2
U^{}_\delta\hat{O}^{}_2\hat{O}^{}_1\hat{O}^{}_3$. After a phase
transformation, one can derive $\tilde{V}=U'\tilde{V}' =U'O^{}_2
U^{}_\delta\hat{O}^{}_2\hat{O}^{}_1\hat{O}^{}_3$ with an additional
unphysical phase matrix $U'$. Comparing it with Eq. \eqref{Vt}, we
find the following relations:
\begin{eqnarray}\label{tildehat}
\tilde{\theta}^{}_1=\hat{\theta}^{}_1\;,&
\tilde{\theta}^{}_3=\hat{\theta}^{}_3\;, & \tilde{s}^2_2=c^2_2
\hat{s}^2_2+s^2_2\hat{c}^2_2
+2c^{}_2s^{}_2\hat{c}^{}_2\hat{s}^{}_2\cos\delta^{}_\text{KM}\;,
\end{eqnarray}
and
\begin{eqnarray}\label{Toshev}
\sin\tilde{\delta}_\text{KM}\sin2\tilde{\theta}^{}_2=\sin\delta^{}_\text{KM}\sin2\theta^{}_2\;,
\end{eqnarray}
where $\hat{s}^{}_2\equiv\sin\hat{\theta}^{}_2$, and
$\hat{c}^{}_2\equiv\cos\hat{\theta}^{}_2$.

Eq. \eqref{Toshev} gives a simple relationship between the genuine
and matter-corrected flavor mixing parameters, which is similar to
the Toshev relation \cite{Toshev} in the PDG parametrization
\begin{eqnarray}
\sin\tilde{\delta}\sin2\tilde{\theta}^{}_{23}
=\sin\delta\sin2\theta^{}_{23}\;,
\end{eqnarray}
in which the tildes always stand for the parameters in matter. We
point out that such a similar relationship exists in not only  the
PDG parametrization, but also the KM parametrization and another
parametrization denoted as P7 in Ref. \cite{FXzz}
\footnote{The P7 parametrization of the MNSP matrix is given by
\begin{eqnarray}
V=\left(
\begin{array}{ccc}
 c'_{12} c'_{13} & s'_{12} & -c'_{12} s'_{13} \\
 -s'_{12} c'_{13} c'_{23}+ s'_{13} s'_{23}e^{-i \delta '} & c'_{12} c'_{23} & s'_{12} s'_{13} c'_{23}+ c'_{13} s'_{23}e^{-i \delta '} \\
  s'_{12} c'_{13} s'_{23}+ s'_{13} c'_{23}e^{-i \delta '} & -c'_{12} s'_{23} & -s'_{12} s'_{13} s'_{23}+ c'_{13} c'_{23}e^{-i \delta '}
\end{array}
\right)\;,\nonumber
\end{eqnarray}
where $s'_{ij}=\sin\theta'_{ij}$ and $c'_{ij}=\cos\theta'_{ij}$. In
this parametrization, the Toshev relation is expressed as
$\sin\tilde{\delta}'\sin2\tilde{\theta}'_{23}
=\sin\delta'\sin2\theta'_{23}$. The reason for the Toshev relation
being satisfied is that the first rotation matrix on the right-hand
side of the MNSP matrix is a (2,3) rotation matrix in all of these
three parametrizations, and it commutes with the effective
potential.}.
Generally, we call all of them the Toshev relations.

We end this part with a direct application of this relation. Given
the $\mu$-$\tau$ symmetry with $\theta^{}_2=45^{\circ}_{}$ and
maximal CP violation with $\delta^{}_\text{KM}=\pm90^{\circ}_{}$ in
vacuum, both sides of the Toshev relation equal $\pm1$, then
$|\sin\tilde{\delta}_\text{KM}|=\;|\sin2\tilde{\theta}^{}_2|=1$,
leading to $\tilde{\theta}^{}_2=45^{\circ}_{}$ and
$\tilde{\delta}_\text{KM}=\pm90^{\circ}_{}$. Thus, we have proved
that the $\mu$-$\tau$ symmetry and maximal CP violation keep
unchanged when matter effects are taken into account. Together with
the PDG and P7 parametrizations, we rewrite the overall invariance
of the $\mu$-$\tau$ symmetry and maximal CP violation in three
parametrizations as follows:
\begin{eqnarray}
&&\theta^{}_{23}=45^{\circ}_{},\delta=\pm90^{\circ}_{}
\Longleftrightarrow
\tilde{\theta}^{}_{23}=45^{\circ}_{},\tilde{\delta}=\pm90^{\circ}_{} \nonumber\\
&&
\Longleftrightarrow\theta^{}_2=45^{\circ}_{},\delta^{}_\text{KM}=\pm90^{\circ}_{}
\Longleftrightarrow
\tilde{\theta}^{}_2=45^{\circ}_{},\tilde{\delta}^{}_\text{KM}=\pm90^{\circ}_{}\;\nonumber\\
&&\Longleftrightarrow\theta'_{23}=45^{\circ}_{},\delta'=\pm90^{\circ}_{}
\Longleftrightarrow
\tilde{\theta}'_{23}=45^{\circ}_{},\tilde{\delta}'=\pm90^{\circ}_{}\;.\label{equal}
\end{eqnarray}
Ref. \cite{XZ} provides another proof to this claim in the PDG
parametrization, where a special basis of the neutrino fields is
taken.

\subsection{Analytical approximations}

Our approximate formulation is based on two premises:
\begin{itemize}
\item From Table 1, one can derive $s^2_3\approx0.04$ (0.05) for the NH
(IH). So $s^2_3\sim |\alpha| \approx0.03$ provides a reliable basis
for our analytical approximation.

\item Given small $\alpha$, two scenarios should be considered separately: (a) $E$
is above $\mathcal{O}(1)$ GeV; and (b) $E$ is below $\mathcal{O}(1)$
GeV.
\end{itemize}
The reason for distinguishing between scenarios (a) and (b) is
simple. In scenario (b), where $|\alpha|\gtrsim |A|$ holds, we have
to regard $A$ as a small parameter; but in scenario (a) with
$|\alpha|\ll |A|$, we do not have to do so.

\subsubsection*{Scenario (a)}

Our first step is to diagonalize $M$ in Eq. \eqref{H1}. The
symmetric matrix $M$ can be decomposed into two terms in series of
$\alpha$:
\begin{eqnarray}\label{Malpha}
&M=M^{(0)}_{}+\alpha M^{(1)}_{}\;.
\end{eqnarray}
The first term $M^{(0)}_{}$ is a singular matrix and can be strictly
diagonalized. The second term $\alpha M^{(1)}_{}$ is a perturbation
and can be diagonalized in series of $\alpha$ order by order. In
general, the eigenvalues and eigenvectors of $M$ are expressed as
\begin{eqnarray}\label{lam}
&\lambda^{}_i=\lambda^{(0)}_i+\alpha\lambda^{(1)}_i+\cdots\;,\nonumber\\
&v^{}_i=v^{(0)}_i+\alpha v^{(1)}_i+\cdots\;,
\end{eqnarray}
respectively. And the orthogonal matrix $\hat{V}$ is written as
$\hat{V}=\left(v^{}_1, v^{}_2, v^{}_3\right)$. Here one must pay
attention to the detail of how to derive a proper order of
$\lambda^{}_i$ (or $v^{}_i$). In order to derive the proper order,
we have to consider the cases of $A<1$ and $A>1$ separately. For
$A<1$, we formulate the series expansion in $\alpha$ and derive the
eigenvalues and eigenvectors to the first order. The eigenvalues are
given by
\begin{eqnarray}\label{lamda}
&&\lambda_1 = \alpha\frac{c^2_1}{1-s^2_1s^2_3}\;,\nonumber\\
&&\lambda_2 = \frac{1}{2}\left[1+ A - C +\alpha s^2_1c^2_3 \frac{ C
+\left(1- A +2 A s^2_1s^2_3\right)}
{ C (1-s^2_1s^2_3)}\right]\;,\nonumber\\
&&\lambda_3 = \frac{1}{2}\left[1+ A + C +\alpha s^2_1c^2_3\frac{ C
-\left(1- A +2 A s^2_1s^2_3\right)} { C (1-s^2_1s^2_3)}\right]\;
\end{eqnarray}
with
\begin{eqnarray}\label{C}
C =\sqrt{(1- A )^2+4 A s^2_1s^2_3}\;.
\end{eqnarray}
The eigenvectors are too complicated to be listed here.

Up to $\mathcal{O}(s^2_3)$, Eq. \eqref{lamda} can be simplified to
\begin{eqnarray}\label{lambda}
&&\lambda_1 = \alpha c^2_1\;,\nonumber\\
&&\lambda_2 = A -\frac{ As^2_1s^2_3 }{1- A }+\alpha s^2_1\;,\nonumber\\
&&\lambda_3 = 1 +\frac{ As^2_1s^2_3 }{1- A }\;.
\end{eqnarray}
The orthogonal matrix $\hat{V}$ turns out to be
\begin{eqnarray}
\hat{V} =\begin{pmatrix}\displaystyle\alpha\frac{c^{}_1s^{}_1}{A}
&\displaystyle-1+\frac{s^2_1s^2_3}{2(1-A)^2}&\displaystyle-\frac{s^{}_1s^{}_3}{1-A}\\
\displaystyle1-\frac{1}{2}c^2_1s^2_3& \displaystyle\alpha\frac{c^{}_1s^{}_1}{A}-\frac{c^{}_1s^{}_1s^2_3}{1-A} &c^{}_1s^{}_3\\
\displaystyle-c^{}_1s^{}_3&\displaystyle-\frac{s^{}_1s^{}_3}{1-A}&\displaystyle1-\frac{s^2_1s^2_3}{2(1-
A )^2}-\frac{1}{2}c^2_1s^2_3
\end{pmatrix}\;,
\end{eqnarray}
whose three rotation angles are given by
$\cot\hat{\theta}^{}_1=\alpha c^{}_1s^{}_1/A$, $\displaystyle
\tan\hat{\theta}^{}_2=-c^{}_1s^{}_3$ and
$\displaystyle\tan\hat{\theta}^{}_3=s^{}_1s^{}_3/(1- A )$.
Substituting the angles into Eqs. \eqref{tildehat} and
\eqref{Toshev}, we obtain the approximate expressions of the
matter-corrected flavor mixing parameters:
\begin{eqnarray}\label{para1}
&&\cot\tilde{\theta}^{}_1=\frac{\alpha\sin2\theta^{}_1}{2A}\;,\nonumber\\
&&\sin\tilde{\theta}^{}_2=\sin\theta^{}_2\sqrt{
\frac{1-2\epsilon\cos\delta^{}_\text{KM}\cot\theta^{}_2+\epsilon^2_{}\cot^2_{}\theta^{}_2}
{1+\epsilon^2_{}}} \;,\nonumber\\
&&\tan\tilde{\theta}^{}_3=\frac{\sin\theta^{}_1\sin\theta^{}_3}{1-A}\;,\nonumber\\
&&\sin\tilde{\delta}_\text{KM}=
\frac{\left(1+\epsilon^2_{}\right)\sin\delta^{}_\text{KM}}
{\sqrt{\left(1+2\epsilon\cos\delta^{}_\text{KM}\tan\theta^{}_2+\epsilon^2_{}\tan^2_{}\theta^{}_2\right)
\left(1-2\epsilon\cos\delta^{}_\text{KM}\cot\theta^{}_2+\epsilon^2_{}\cot^2_{}\theta^{}_2\right)}}\;,
\end{eqnarray}
in which $\epsilon=c^{}_1s^{}_3$. One can see that
$\cot\tilde{\theta}^{}_1$ and $\tan\tilde{\theta}^{}_3$ are
suppressed by $\alpha$ or $s^{}_3$, leading to
$\tilde{\theta}^{}_1\sim 90^{\circ}$ and $\tilde{\theta}^{}_3\sim
0$. The most interesting result comes from $\tilde{\theta}^{}_2$ and
$\tilde{\delta}^{}_\text{KM}$. First,
$\tilde{\theta}^{}_2\approx\theta^{}_2$ and
$\tilde{\delta}^{}_\text{KM}\approx\delta^{}_\text{KM}$ are good
approximations for $\epsilon$ is suppressed by $s^{}_3$. In other
words, the matter effects on $\tilde{\theta}^{}_2$ and
$\tilde{\delta}^{}_\text{KM}$ are small. Second, such small matter
effects are independent of $A$ in the approximation made above. This
is a peculiar feature of the KM parametrization of $\tilde{V}$.

Note that Eq. \eqref{para1} only holds for neutrinos with $A<1$. For
neutrinos with $A>1$, the replacement of
$\tan\tilde{\theta}^{}_3\Rightarrow\cot\tilde{\theta}^{}_3$ in Eq.
\eqref{para1} should be made.

For antineutrinos with $-A<1$, the approximate expressions are given
by
\begin{eqnarray}\label{para2}
&&\tan\tilde{\theta}^{}_1=\frac{\sin\theta^{}_1\sin\theta^{}_3}{1+ A}\;,\nonumber\\
&&\sin\tilde{\theta}^{}_2=\cos\theta^{}_2\sqrt{
\frac{1+2\varepsilon\cos\delta^{}_\text{KM}\tan\theta^{}_2+\varepsilon^2_{}\tan^2_{}\theta^{}_2}
{1+\varepsilon^2_{}}}\;,\nonumber\\
&&\cot\tilde{\theta}^{}_3=\frac{\alpha(1+ A)\cos\theta^{}_1}{ A \sin\theta^{}_3}\;,\nonumber\\
&&\sin\tilde{\delta}_\text{KM}=
\frac{\left(1+\varepsilon^2_{}\right)\sin\delta^{}_\text{KM}}
{\sqrt{\left(1+2\varepsilon\cos\delta^{}_\text{KM}\tan\theta^{}_2+\varepsilon^2_{}\tan^2_{}\theta^{}_2\right)
\left(1-2\varepsilon\cos\delta^{}_\text{KM}\cot\theta^{}_2+\varepsilon^2_{}\cot^2_{}\theta^{}_2\right)}}\;,
\end{eqnarray}
in which $\displaystyle
\varepsilon=c^{}_1s^{}_3[1-\alpha(1+A)/(As^2_3)]$. Eq. \eqref{para2}
is different from Eq. \eqref{para1} because the order of
$\lambda^{}_1$ and $\lambda^{}_2$ has been exchanged. Given small
$\theta^{}_3$, $\tilde{\theta}^{}_1\sim0$,
$\tilde{\theta}^{}_3\sim90^\circ_{}$ and
$\tilde{\theta}^{}_2\sim90^\circ_{}-\theta^{}_2$ roughly hold. As
$\varepsilon$ is $A$-dependent, $\tilde{\theta}^{}_2$ and
$\tilde{\delta}^{}_\text{KM}$ obviously rely on $A$.

For antineutrinos with $-A>1$, the replacement of
$\tan\tilde{\theta}^{}_1\Rightarrow\cot\tilde{\theta}^{}_1$ in Eq.
\eqref{para2} should be made.

The magnitude of the intrinsic CP violation in neutrino oscillations
depends only upon the Jarlskog invariant \cite{Jarl}. It is
calculated via $\mathcal{J}=\text{Im} \left(V^{}_{e1}V^{}_{\mu2}
V^*_{e2}V^*_{\mu1}\right)$. In the KM parametrization, one has
$\mathcal{J}= (1/8)\sin\theta^{}_1 \sin2\theta^{}_1 \sin2\theta^{}_2
\sin2\theta^{}_3 \sin \delta^{}_{\text{KM}}$ in vacuum and
$\tilde{\mathcal{J}}= (1/8)
\sin\tilde{\theta}^{}_1\sin2\tilde{\theta}^{}_1
\sin2\tilde{\theta}^{}_2 \sin2\tilde{\theta}^{}_3\sin
\tilde{\delta}^{}_{\text{KM}}$ in matter. They satisfy the Naumov
identity $\mathcal{\tilde{J}} \Delta \tilde{m}^2_{21}\Delta
\tilde{m}^2_{31} \Delta \tilde{m}^2_{32} = \mathcal{J} \Delta
m^2_{21} \Delta m^2_{31} \Delta m^2_{32} $  \cite{Naumov,Xzz}. One
can derive $\mathcal{\tilde{J}}$ from Eq. \eqref{lamda}  as
\begin{eqnarray}\label{Jarlskog1}
\mathcal{\tilde{J}}=\frac{\alpha}{AC
\left(1-\sin^2_{}\theta^{}_1\sin^2_{}\theta^{}_3\right)}\mathcal{J}
\end{eqnarray}
for neutrinos. Suppressed by $\alpha$, the Jarlskog invariant in
matter is much smaller than that in vacuum. $\mathcal{\tilde{J}}$
obtains its relative maximum $\mathcal{\tilde{J}}^\text{a}_\text{m}
\approx\alpha\mathcal{J}/(2 s^{}_1 s^{}_3)
=(1/8)\alpha\sin2\theta^{}_1\sin2\theta^{}_2\cos\theta^{}_3
\sin\delta^{}_\text{KM}$ at $A\approx 1$, where $C$ is near its
minimal value. The series expansion in $s^{}_3$ gives the
leading-order result of $\mathcal{\tilde{J}}$
\begin{eqnarray}\label{Jarlskog2}
\mathcal{\tilde{J}}=\frac{\alpha}{ A (1- A )}\mathcal{J}\;.
\end{eqnarray}
It can also be obtained from Eq. \eqref{para1}. Eq.
\eqref{Jarlskog2} does not hold for $A\sim 1$. To derive the correct
result of $\mathcal{\tilde{J}}$ for antineutrinos, one has to make
the replacements $A\Rightarrow-A$ and
$\mathcal{J}\Rightarrow-\mathcal{J}$.

\subsubsection*{Scenario (b)}

Considering about $|A|\lesssim|\alpha|$, we rewrite Eq.
\eqref{Malpha} as
\begin{eqnarray}\label{MalphaA}
M={M'}^{(0)}_{}+A {M'}^{(1)}_{}+\alpha M^{(1)}_{}\;,
\end{eqnarray}
where ${M'}^{(0)}_{}+A {M'}^{(1)}_{}\equiv{M}^{(0)}_{}$.
${M'}^{(0)}_{}$ has two degenerate eigenvalues 0 and one eigenvalue
1. To the first order of $\alpha$, the eigenvalues are expressed as
\begin{eqnarray}\label{}
&&\lambda^{}_1=\frac{1}{2}\left[A(1-s^2_1s^2_3)+\alpha-D\right]\;,\nonumber\\
&&\lambda^{}_2=\frac{1}{2}\left[A(1-s^2_1s^2_3)+\alpha+D\right]\;,\nonumber\\
&&\lambda^{}_3=1+As^{}_1c^{}_3s^{}_3\;,
\end{eqnarray}
in which
$D=\sqrt{\left[A\left(1-s^2_1s^2_3\right)+\alpha\right]^2_{}-4\alpha
Ac^2_1}$. $M$ is diagonalized by
\begin{eqnarray}\label{}
\hat{V}=O^{}_1O^{}_3\left(
\begin{array}{ccc}
 \cos\vartheta & -\sin\vartheta & 0 \\
 \sin\vartheta & \cos\vartheta & 0 \\
 0 & 0 & 1
\end{array}
\right)\left(
\begin{array}{ccc}
 1 & 0 & h^{}_{13} \\
 0 & 1 & h^{}_{23} \\
 -h^{}_{13} & -h^{}_{23} & 1
\end{array}
\right)\;
\end{eqnarray}
with one possibly large rotation angle $\vartheta$ at the zeroth
order in $A$ and two small perturbation coefficients $h^{}_{13}$ and
$h^{}_{23}$ at the first order. A strict calculation yields
\begin{eqnarray}\label{vartheta}
&&\sin\vartheta=\sqrt{\frac{1}{2}+\frac{A\cos2\theta^{}_1-\alpha}{2D}}\;,\nonumber\\
&&h^{}_{13}=As^{}_1s^{}_3(c^{}_1\sin\vartheta+s^{}_1c^{}_3\cos\vartheta)\;,\nonumber\\
&&h^{}_{23}=As^{}_1s^{}_3(c^{}_1\cos\vartheta-s^{}_1c^{}_3\sin\vartheta)\;.
\end{eqnarray}
Finally, by using Eqs. \eqref{tildehat} and \eqref{Toshev} and
taking account of $s^2_3\sim\alpha$, we obtain the flavor mixing
parameters in matter:
\begin{eqnarray}\label{para3}
&&\sin2\tilde{\theta}^{}_1=\frac{\alpha}{D}\sin2\theta^{}_1\nonumber\;,\\
&&\sin\tilde{\theta}^{}_2=\sin\theta^{}_2\sqrt{
\frac{1-2\kappa\cos\delta^{}_\text{KM}\cot\theta^{}_2+\kappa^2_{}\cot^2_{}\theta^{}_2}{1+\kappa^2_{}}}\;,\nonumber\\
&&\tan\tilde{\theta}^{}_3=\frac{\sin\theta^{}_1\sin\theta^{}_3}{\sin(\theta^{}_1+\vartheta)}\;,\nonumber\\
&&\sin\tilde{\delta}_\text{KM}=\frac{\left(1+\kappa^2_{}\right)\sin\delta^{}_\text{KM}}
{\sqrt{\left(1+2\kappa\cos\delta^{}_\text{KM}\tan\theta^{}_2+\kappa^2_{}\tan^2_{}\theta^{}_2\right)
\left(1-2\kappa\cos\delta^{}_\text{KM}\cot\theta^{}_2+\kappa^2_{}\cot^2_{}\theta^{}_2\right)}}\;,
\end{eqnarray}
where $\kappa=s^{}_3\sin\vartheta\csc(\theta^{}_1+\vartheta)$. If
one sets $A\rightarrow0$, then $D\rightarrow\alpha$,
$\vartheta\rightarrow0$ and all the parameters in matter return to
those in vacuum. For a nonzero $A$, $\tilde{\theta}^{}_1$ may have a
remarkable deviation from $\theta^{}_1$. For example, if
$A=\alpha\cos2\theta^{}_1$, then $D\approx\alpha\sin2\theta^{}_1$
and $\sin2\tilde{\theta}^{}_1\approx1$, leading to
$\tilde{\theta}^{}_1\approx45^\circ_{}$. But if $A=\alpha$, then
$D\approx2\alpha s^{}_1$ and
$\sin2\tilde{\theta}^{}_1\approx\cos\theta^{}_1$, leading to
$\tilde{\theta}^{}_1\approx45^\circ_{}+\theta^{}_1/2$.
$\tilde{\theta}^{}_3$ remains small though it receives some
corrections. $\tilde{\theta}^{}_2$ and $\tilde{\delta}^{}_\text{KM}$
have small $A$-dependent corrections suppressed by $s^{}_3$.

Eq. \eqref{para3} holds for neutrinos with the NH. One can replace
$\vartheta$ with $90^\circ_{}-\vartheta$, $-\vartheta$ and
$\vartheta-90^\circ_{}$ but keep Eq. \eqref{para3} unchanged to
derive the correct results for neutrinos with the IH, antineutrinos
with the NH and antineutrinos with the IH, respectively.

The Jarlskog invariant reads
\begin{eqnarray}\label{Jarlskog3}
\tilde{\mathcal{J}}=\frac{\alpha}{D}\mathcal{J}\;
\end{eqnarray}
for the NH. Since $D\sim A\sim\alpha$, $\tilde{\mathcal{J}}$ is of
the same order as $\mathcal{J}$. Therefore, the CP violation for
$E<\mathcal{O}(1)$ GeV is not suppressed significantly by matter
effects. $\tilde{\mathcal{J}}$ gets its relative maximum
$\tilde{\mathcal{J}}^\text{b}_\text{m} \approx\mathcal{J}\csc
2\theta^{}_1=
(1/8)\sin\theta^{}_1\sin2\theta^{}_2\sin2\theta^{}_3\sin\delta^{}_\text{KM}$
at $A\approx\alpha\cos2\theta^{}_1$, which is even greater than
$\mathcal{J}$. One can obtain the results in the IH case by changing
$\alpha$ to $-\alpha$.

In Table 2 we show how to obtain the expressions of the
matter-induced flavor mixing parameters in all the possible cases we
have discussed in both scenarios.

\subsection{Numerical Analysis}

In Fig. 2 we make a comparison of the analytical and numerical
results for the matter-corrected flavor mixing parameters. The
analytical expressions in different intervals of $A$ are given in
Table 2. We have utilized Eq. \eqref{A} with $\rho^{}_\text{m}=2.8\;
\text{g}/\text{cm}^3_{}$ and replaced $A$ with the neutrino energy
$E$. The values of the other input parameters are taken to be
$\Delta m^2_{21}=7.59\times10^{-5}\;\text{eV}^2_{}$, $\Delta
m^2_{31}=2.5\;(-2.4)\times10^{-3}\;\text{eV}^2_{}$,
$\theta^{}_1=34.5^\circ_{}\;(34.6^\circ_{})$,
$\theta^{}_2=49.3^\circ_{}\;(43.2^\circ_{})$,
$\theta^{}_3=11.6^\circ_{}\;(12.9^\circ_{})$ and
$\delta^{}_\text{KM}=-110^\circ_{}\;(-74^\circ_{})$ for the NH (IH),
according to the best-fit values in Ref. \cite{Schwetz} and Table 1.
Some comments and discussions are in order.
\begin{itemize}

\item In most cases, the analytical results in scenario (a) fit
well with their numerical results for $E>\mathcal{O}(1)$ GeV and the
analytical results in scenario (b) fit well with their numerical
results for $E<\mathcal{O}(1)$ GeV. So $|\alpha|\ll|A|$ corresponds
to $E>\mathcal{O}(1)$ GeV, and $|\alpha|\gtrsim |A|$ corresponds to
$E<\mathcal{O}(1)$ GeV.

\item The expression of $\tilde{\theta}^{}_3$ for neutrinos with the NH
and that of $\tilde{\theta}^{}_1$ for antineutrinos with the IH are
not good approximations in the interval
$9\;\text{GeV}<E<15\;\text{GeV}$. Both $\tilde{\theta}^{}_3$ and
$\tilde{\theta}^{}_1$ increase rapidly from a small angle to near
$90^\circ_{}$ when $E$ is running from 9 GeV to 15 GeV.
$\tilde{\theta}^{}_3$ reaches $45^\circ_{}$ at $ E\approx$ 12 GeV
(or $A\approx1$) for neutrinos with the NH, while
$\tilde{\theta}^{}_1$ reaches $45^\circ_{}$ at the same energy for
antineutrinos with the IH.

\item The analytical approximations of $\tilde{\theta}^{}_2$ and
$\tilde{\delta}^{}_\text{KM}$ are in good agreement with their
numerical results. For neutrinos with $E>\mathcal{O}(1)$ GeV, the
numerical results confirm that $\tilde{\theta}^{}_2$ and
$\tilde{\delta}^{}_\text{KM}$ have small deviations from the vacuum
parameters $\theta^{}_2$ and $\delta^{}_\text{KM}$ and are nearly
independent of the neutrino beam energy $E$.

\end{itemize}

In Fig. 3 we compare the analytical results of the Jarlskog
invariant $\tilde{\mathcal{J}}$ in Eqs. \eqref{Jarlskog1},
\eqref{Jarlskog2} and \eqref{Jarlskog3} with the numerical results.
The input values are taken the same as in Fig. 2. The analytical
approximations fit well with the numerical results, except for Eq.
\eqref{Jarlskog2} in the interval $9\;\text{GeV}<E<15\;\text{GeV}$.
The numerical results show that the relative maximum of the Jarlskog
invariant $\mathcal{J}^\text{a}_\text{m}\approx-0.003$ takes place
at $E\approx$ 12 GeV (or $A\approx1$) for neutrinos with the NH and
antineutrinos with the IH, and
$\mathcal{J}^\text{b}_\text{m}\approx-0.03$ at $E\approx$ 0.1 GeV
(or $A\approx\alpha\cos2\theta^{}_1$) for neutrinos with both the NH
and IH. They verify our analytical calculations of
$\mathcal{J}^\text{a}_\text{m}$ and $\mathcal{J}^\text{b}_\text{m}$
in the above discussions.

\section{The probability of $\nu^{}_\mu\rightarrow\nu^{}_e$ oscillations}

The probability of $\nu^{}_\mu\rightarrow\nu^{}_e$ oscillations in
matter has been calculated in the PDG parametrization
\cite{CFA,Minakata1,Minakata2,Sato} but not yet in the KM
parametrization. In this section, we calculate it in the KM
parametrization. Both scenarios (a) and (b) will be considered.

In vacuum, the oscillation probability of
$\nu^{}_\mu\rightarrow\nu^{}_e$ is given by
\begin{eqnarray}
P\left(\nu^{}_\mu \rightarrow \nu^{}_e\right)&=& -4\sum^3_{i>j}
\text{Re}\left(V_{\mu i}V_{e j}V^*_{\mu  j}V^*_{e i}\right) \sin
\Delta^{}_{ij} -8\mathcal{J} \prod _{i>j}^3 \sin \Delta^{}_{ij}\;,
\end{eqnarray}
in which $\Delta^{}_{ij}\equiv\Delta m^2_{ij} L/(4E)$. In the KM
parametrization, one has
\begin{eqnarray}\label{Pue1}
P\left(\nu^{}_\mu \rightarrow \nu^{}_e\right)&=&
\left(\sin^2_{}\theta^{}_1\sin^2_{}\theta^{}_2\sin^2_{}2\theta^{}_3
+J\sin^2_{}\theta^{}_3\cos\delta^{}_\text{KM}+2\sin^2_{}2\theta^{}_1\cos^2_{}\theta^{}_2\sin^4_{}\theta^{}_3
\right)\sin^2_{}\Delta^{}_{32}\nonumber\\
&&+J\cos\left(\Delta^{}_{31}+\delta^{}_\text{KM}\right)\sin\Delta^{}_{32}\sin\Delta^{}_{21}
+\sin^2_{}2\theta^{}_1\cos\theta^{}_2\cos^2_{}\theta^{}_3\sin^2_{}\Delta^{}_{21}\nonumber\\
&&+2\sin^2_{}2\theta^{}_1\cos^2_{}\theta^{}_2\sin^2_{}\theta^{}_3\cos\Delta^{}_{31}
\sin\Delta^{}_{32}\sin\Delta^{}_{21}\;,
\end{eqnarray}
where $J\equiv8\mathcal{J}/\sin\delta^{}_\text{KM}=\sin\theta^{}_1
\sin2\theta^{}_1 \sin2\theta^{}_2 \sin2\theta^{}_3$. Replacing
$\delta^{}_\text{KM}$ with $-\delta^{}_\text{KM}$, one may obtain
the expression of $P\left(\bar{\nu}^{}_\mu \rightarrow
\bar{\nu}^{}_e\right)$. Their difference is a measure of the
intrinsic CP violation:
\begin{eqnarray}
\Delta P\left(\nu^{}_\mu \rightarrow \nu^{}_e\right)&\equiv&
P\left(\nu _\mu \rightarrow \nu _e\right)
- P\left(\bar{\nu} _\mu \rightarrow \bar{\nu} _e\right)\nonumber\\
&=&-2J\sin\delta^{}_\text{KM}\sin\Delta^{}_{31}\sin\Delta^{}_{32}\sin\Delta^{}_{21}\;.
\end{eqnarray}

By replacing the parameters in vacuum with those in matter, one can
achieve a similar expression of the oscillation probability in
matter $\tilde{P}\left(\nu^{}_\mu \rightarrow \nu^{}_e\right)$ from
Eq. \eqref{Pue1}. In scenario (a), with the help of Eq.
\eqref{para1}, $\tilde{P}\left(\nu^{}_\mu\rightarrow\nu^{}_e\right)$
will be re-expressed in terms of the genuine mixing parameters and
$A$. It is not difficult to derive the following approximate formula
from Eq. \eqref{Pue1} by replacing $\Delta^{}_{32}$ and
$\Delta^{}_{21}$ with
$\displaystyle\frac{\sin(1-A)\Delta^{}_{31}}{1-A}$ and
$\displaystyle\alpha\frac{\sin A\Delta^{}_{31} }{ A}$, respectively:
\begin{eqnarray}\label{Pue2}
\tilde{P}\left(\nu^{}_\mu\rightarrow\nu^{}_e\right) &=&
\left(\sin^2_{}\theta^{}_1\sin^2_{} \theta^{}_2 \sin^2_{}
2\theta^{}_3 +
  J\sin^2_{}\theta^{}_3 \cos  \delta
_{\text{KM}}\right)\frac{\sin ^2(1-A)\Delta^{}_{31}
}{(1-A)^2}\nonumber\\
&&+\alpha J \cos \left( \Delta^{}_{31}
+\delta^{}_{\text{KM}}\right)\frac{\sin A
\Delta^{}_{31} }{ A}\frac{\sin (1-A)\Delta^{}_{31} }{1-A}\nonumber\\
&&+\alpha^2_{}\sin^2_{}2\theta^{}_1\cos^2_{}\theta^{}_2\cos^2_{}\theta^{}_3\frac{\sin^2_{}A\Delta^{}_{31}}{A^2_{}}\;.
\end{eqnarray}
Here $\Delta^{}_{31}=\tilde{\Delta}^{}_{31}$,
$(1-A)\Delta^{}_{31}=\tilde{\Delta}^{}_{32}$ and $
A\Delta^{}_{31}=\tilde{\Delta}^{}_{21}$ hold to the leading order in
$\alpha$ and $s^{}_3$, and Eq. \eqref{Pue2} holds to the second
order in $\alpha$ and the third order in $s^{}_3$. The second term
in the first line is a term of $s^3_3$. We keep it in view of the
sizable value of $\theta^{}_3$. The CP violation is included in the
second line, which is suppressed by $\alpha$. The term of
$\alpha^2_{}$ in the third line is reserved for comparing with some
former works in the PDG parametrization, in which the term of
$\alpha^2_{}$ is often kept ({\it e.g.}, \cite{CFA}). But given the
current experimental data, this term is meaningless unless we keep
the terms of $s^4_{13}$ or $\alpha s^2_{13}$ \cite{Minakata1}. The
divergences for $A\rightarrow0$ and $1$ are absent because of the
suppressions by sine functions $\sin \hat{A}\Delta$ and $\sin
(\hat{A}-1)\Delta$. One can see from Eq. \eqref{Pue1} that the
expression of the $\nu^{}_\mu\rightarrow\nu^{}_e$ oscillation
probability in the KM parametrization is similar to that in the PDG
parametrization. By replacing $\delta^{}_\text{KM}$ with
$-\delta^{}_\text{KM}$ and $A$ with $-A$ in Eq. \eqref{Pue2}, one
can obtain the expression of
$\tilde{P}(\bar{\nu}^{}_\mu\rightarrow\bar{\nu}^{}_e)$ and then
their difference $\Delta\tilde{P}\left(\nu^{}_\mu \rightarrow
\nu^{}_e\right) \equiv \tilde{P}(\nu^{}_\mu\rightarrow\nu^{}_e) -
\tilde{P}(\bar{\nu}^{}_\mu\rightarrow\bar{\nu}^{}_e)$.
$\Delta\tilde{P}\left(\nu^{}_\mu \rightarrow \nu^{}_e\right)$
consists of both the intrinsic CP violation and matter-induced
contribution and is sometimes dominated by the latter.

In scenario (b), one can also make the similar replacements as in
scenario (a) and obtain the approximate expression of
$\tilde{P}(\nu^{}_\mu\rightarrow\nu^{}_e)$ in terms of the vacuum
parameters and $A$. But because of the complexity of $\vartheta$, it
is difficult for us to simplify the expression to a form as a simple
function of $A$ \cite{Minakata2}. So we do not write out the
expression here. We use another method proposed in Ref. \cite{Sato}
to obtain an alternative simple formula. It reads
\begin{eqnarray}
\tilde{P}\left(\nu^{}_\mu \rightarrow \nu^{}_e\right)
&=&P\left(\nu^{}_\mu \rightarrow \nu^{}_e\right)
+2A\sin\Delta^{}_{31}\left(\sin\Delta^{}_{31} -\Delta^{}_{31}
\cos\Delta^{}_{32}\right)\nonumber\\
&&\times\left(\sin ^2\theta^{}_1\sin ^2 \theta^{}_2 \sin ^2
2\theta^{}_3 + J\sin^2_{}\theta^{}_3 \cos  \delta
_{\text{KM}}\right)\;.
\end{eqnarray}
With $\delta^{}_\text{KM}\Rightarrow-\delta^{}_\text{KM}$ and
$A\Rightarrow-A$, one may arrive at $\tilde{P}\left(\bar{\nu}^{}_\mu
 \rightarrow \bar{\nu}^{}_e\right)$. Their difference is given by
\begin{eqnarray}\label{PueD}
\Delta\tilde{P}\left(\nu^{}_\mu \rightarrow \nu^{}_e\right)
&=&\Delta P\left(\nu^{}_{\mu } \rightarrow \nu^{}_e\right)
+4A\sin\Delta^{}_{31}\left(\sin\Delta^{}_{31}-\Delta^{}_{31}\cos\Delta^{}_{32}\right)\nonumber\\
&&\times\left(\sin ^2\theta^{}_1\sin ^2 \theta^{}_2 \sin ^2
2\theta^{}_3 + J\sin^2_{}\theta^{}_3 \cos  \delta
_{\text{KM}}\right)\;.
\end{eqnarray}
In Eq. \eqref{PueD}, the intrinsic CP violation and the
matter-induced contribution are separated to two different parts.
The former is dominant if $\delta^{}_\text{KM}$ is not too small.

Finally, we compare the analytical results of
$\tilde{P}\left(\nu^{}_\mu\rightarrow\nu^{}_e\right)$ and
$\Delta\tilde{P}\left(\nu^{}_\mu\rightarrow\nu^{}_e\right)$ with the
numerical results to show the validity of our approximations. We
take the future long-baseline neutrino oscillation experiments
NO$\nu$A as an example in scenario (a) and Hyper-K as an example in
scenario (b). The former will use the neutrino beams with $E\sim$ 2
GeV, which is compatible with scenario (a). And the latter will use
the neutrino beams with $E\sim$ 0.6 GeV, which is compatible with
scenario (b). Different baseline lengths $L$ and matter densities
$\rho^{}_\text{m}$ have been input in different experiments. $L=810$
km and $\rho^{}_\text{m}=2.8\;\text{g}/\text{cm}^3_{}$ are taken in
Fig. 4 for the NO$\nu$A experiment \cite{NOvA}, and $L=295$ km and
$\rho^{}_\text{m}=2.6\;\text{g}/\text{cm}^3_{}$ are taken in Fig. 5
for the Hyper-K experiment \cite{HyperK}. Other input parameters are
taken the same as in Fig. 2. We see that the numerical results
confirm the precision of the analytical approximations of
$\tilde{P}\left(\nu^{}_\mu\rightarrow\nu^{}_e\right)$ and
$\Delta\tilde{P}\left(\nu^{}_\mu\rightarrow\nu^{}_e\right)$ in both
scenarios.

\section{Conclusions}
In this paper, we have used the KM parametrization to study the
lepton flavor mixing and neutrino oscillations in matter. We
re-discover the Toshev-like relation in the KM parametrization and
prove that the $\mu$-$\tau$ symmetry with $\theta^{}_2=45^\circ_{}$
and maximal CP violation with $\delta^{}_\text{KM}=\pm90^\circ_{}$
keep unchanged when matter effects are taken into account. We have
presented the approximate expressions of the matter-corrected flavor
mixing parameters. Different methods have been chosen in two
scenarios for the neutrino energy above $\mathcal{O}(1)$ GeV and
below $\mathcal{O}(1)$ GeV. Finally, we have calculated the
probability of $\nu^{}_\mu\rightarrow\nu^{}_e$ oscillations as an
application in both scenarios. Below we compare the main features of
the KM parametrization with those of the PDG parametrization:
\begin{itemize}
\item The genuine flavor mixing parameters in the KM parametrization
$\theta^{}_1$, $\theta^{}_2$ and $\delta^{}_\text{KM}$ are
approximately equal to the corresponding parameters in the PDG
parametrization $\theta^{}_{12}$, $\theta^{}_{23}$ and $\delta$,
respectively. Although $\theta^{}_3$ is not close to
$\theta^{}_{13}$, it is small enough to do the series expansion in
both $\theta^{}_3$ and $\Delta m^2_{21}/\Delta m^2_{31}$ by treating
$\sin^2_{}\theta^{}_3\sim\Delta m^2_{21}/\Delta m^2_{31}$.

\item  For neutrinos with $E$ above $\mathcal{O}(1)$ GeV, the
corrections to the rotation angle $\tilde{\theta}^{}_2$ and the CP
phase $\tilde{\delta}^{}_\text{KM}$ induced by matter effects are
small and nearly independent of the matter density and the neutrino
energy. This is a salient feature of the KM parametrization.

\item The analytical expressions of the oscillation
probabilities in Eqs. \eqref{Pue1} and \eqref{Pue2} are similar to
the corresponding expressions in the PDG parametrization.
\end{itemize}

It is well known that a good parametrization brings much convenience
to the description of physical quantities. In this paper, we have
explored the KM parametrization to study the neutrino phenomenology.

\vspace{0.5cm}
\section*{Acknowledgments}
The author would like to thank Prof. Z.Z. Xing for suggesting this
work and correcting the manuscript in great detail. He is also
grateful to Y.F. Li for useful discussions. This work was supported
in part by the National Natural Science Foundation of China under
grant No. 10875131.

\vspace{0.5cm}

\begin{table}[!h]
\centering
\begin{tabular}{cccc}
\hline\hline
 Parameter & $\theta^{}_1$
 & $\theta^{}_2$ ($\delta$ at its best fit) & $\theta^{}_3$ \\
\hline\hline
 Best fit       & $34.5^\circ_{}$                      & $49.3^\circ_{}$           & $11.6^\circ_{}$                                 \\
                & ($34.6^\circ_{}$)                    & ($43.2^\circ_{}$)         & ($12.9^\circ_{}$)                               \\\hline
 1$\sigma$ range& $33.4^\circ_{}$ - $35.8^\circ_{}$   & $44.8^\circ_{}$ - $53.0^\circ_{}$   & $9.4^\circ_{}$ - $14.0^\circ_{}$    \\
                & ($33.5^\circ_{}$ - $36.0^\circ_{}$) & ($39.8^\circ_{}$ - $45.8^\circ_{}$) & ($10.4^\circ_{}$ - $15.3^\circ_{}$) \\\hline
 2$\sigma$ range& $32.1^\circ_{}$ - $37.4^\circ_{}$   & $41.9^\circ_{}$ - $55.0^\circ_{}$   & $6.8^\circ_{}$ - $16.0^\circ_{}$    \\
                & ($32.2^\circ_{}$ - $37.5^\circ_{}$) & ($38.4^\circ_{}$ - $47.1^\circ_{}$) & ($7.6^\circ_{}$ - $16.8^\circ_{}$)  \\\hline
 3$\sigma$ range& $31.4^\circ_{}$ - $38.2^\circ_{}$   & $39.7^\circ_{}$ - $56.8^\circ_{}$   & $3.5^\circ_{}$ - $17.6^\circ_{}$    \\
                & ($31.4^\circ_{}$ - $38.4^\circ_{}$) & ($37.8^\circ_{}$ - $48.2^\circ_{}$) & ($3.5^\circ_{}$ - $18.6^\circ_{}$)  \\
\hline\hline
\end{tabular}\label{S}
 \caption{The mixing angles in the KM parametrization
translated from the results obtained in the PDG parametrization
\cite{Schwetz}. The upper (lower) row corresponds to the normal
hierarchy (inverted hierarchy) of neutrino masses. The best-fit
value $\delta=-110^\circ_{}$ ($-74^\circ_{}$) have been taken  for
simplicity.}
 \vspace{0.3cm}
\end{table}

\begin{table}[!h]
\centering
\begin{tabular}{l|ll|ll}
\hline\hline
   &\;NH& &\;IH& \\\hline
$\nu$   & $A\lesssim\alpha$ &Eq. \eqref{para3} & $|A|\lesssim|\alpha|$ &Eq. \eqref{para3}, $\vartheta\Rightarrow90^\circ_{}-\vartheta$ \\
   & $\alpha\ll A<1$      &Eq. \eqref{para1} & $|\alpha|\ll|A|$ &  Eq. \eqref{para1} \\
   & $1<A$             &Eq. \eqref{para1}, $\theta^{}_3\Rightarrow90^\circ_{}-\theta^{}_3 $&&\\\hline
$\bar{\nu}$   & $A\lesssim\alpha$ & Eq. \eqref{para3}, $\vartheta\Rightarrow-\vartheta$ & $|A|\lesssim|\alpha|$ & Eq. \eqref{para3}, $\vartheta\Rightarrow\vartheta-90^\circ_{}$  \\
   & $\alpha\ll A$ & Eq. \eqref{para2} & $|\alpha|\ll|A|<1$ & Eq. \eqref{para2} \\
   &   &   & $1<|A|$ & Eq. \eqref{para2}, $\theta^{}_3\Rightarrow90^\circ_{}-\theta^{}_3 $  \\\hline\hline
\end{tabular}\label{S}
\caption{The replacements for obtaining the proper expressions of
the effective matter-corrected flavor mixing parameters in different
cases.}
 \vspace{0.3cm}
\end{table}

\vspace{0.5cm}

\begin{figure}[!h]
\center\label{t2}
\begin{overpic}[width=11cm]{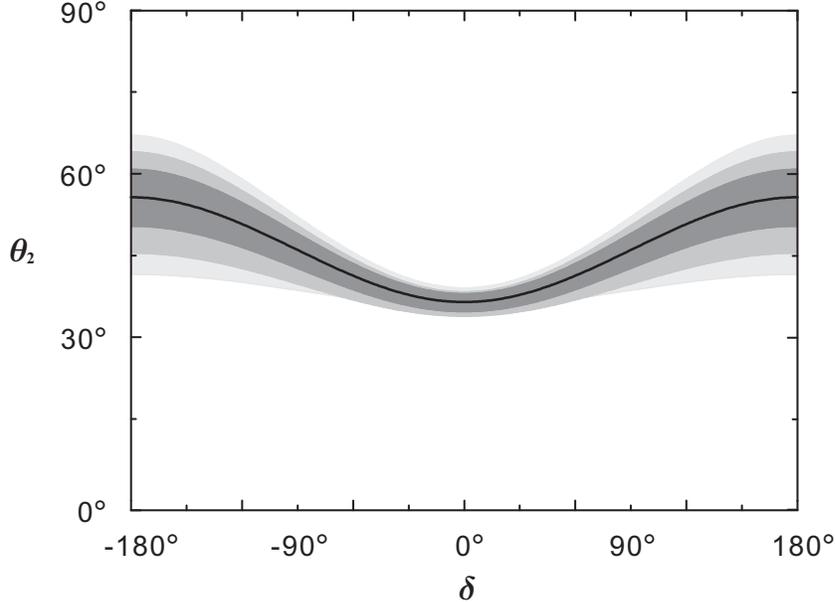}
\end{overpic}\caption{The dependence of $\theta^{}_2$ on $\delta$
according to Eq. \eqref{kmpdg}, where the best-fit values and the
$1\sigma$, $2\sigma$ and $3\sigma$ ranges of $\theta^{}_{12}$,
$\theta^{}_{13}$ and $\theta^{}_{23}$ \cite{Schwetz} have been
input.}
\end{figure}

\begin{figure}[!h]
\center\label{mf}
\begin{overpic}[width=15cm]{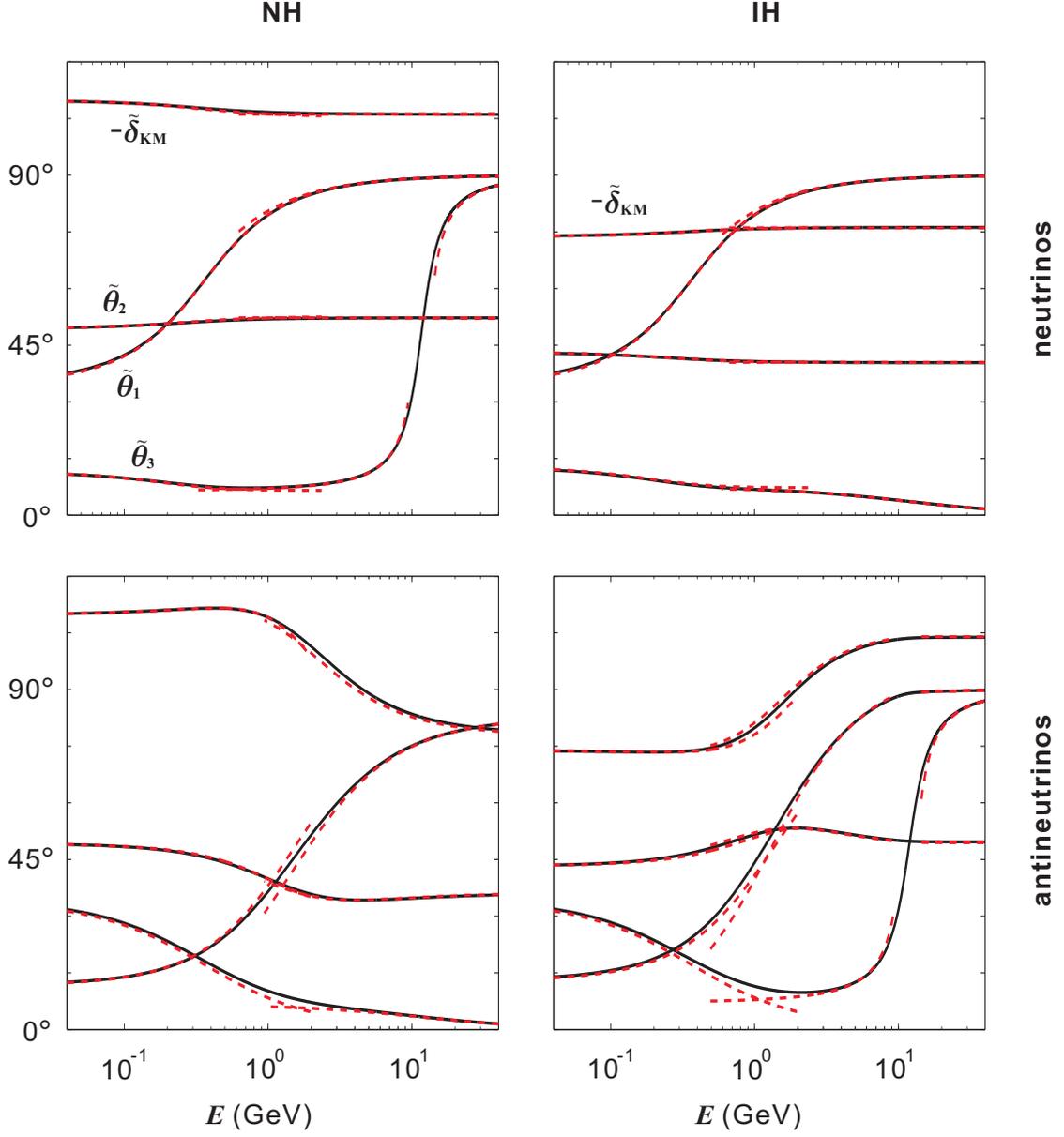}
\end{overpic}\caption{The analytical (dashed line)
and numerical (solid line) results of the matter-corrected flavor
mixing parameters. The genuine flavor mixing parameters are taken as
the best-fit values in Ref. \cite{Schwetz} and Table 1: $\Delta
m^2_{21}=7.59\times10^{-5}\;\text{eV}^2_{}$, $\Delta
m^2_{31}=2.5\;(-2.4)\;\times10^{-3}\;\text{eV}^2_{}$,
$\theta^{}_1=34.5^\circ_{}\;(34.6^\circ_{})$,
$\theta^{}_3=11.6^\circ_{}\;(12.9^\circ_{})$,
$\theta^{}_2=49.3^\circ_{} (43.2^\circ_{})$ and
$\delta^{}_\text{KM}=-110^\circ_{}\;(-74^\circ_{})$ for the NH (IH).
The mass density of matter is assumed to be
$\rho^{}_\text{m}=2.8\;\text{g}/\text{cm}^3$.}
\end{figure}

\begin{figure}[!h]
\center\label{mf}
\begin{overpic}[width=15cm]{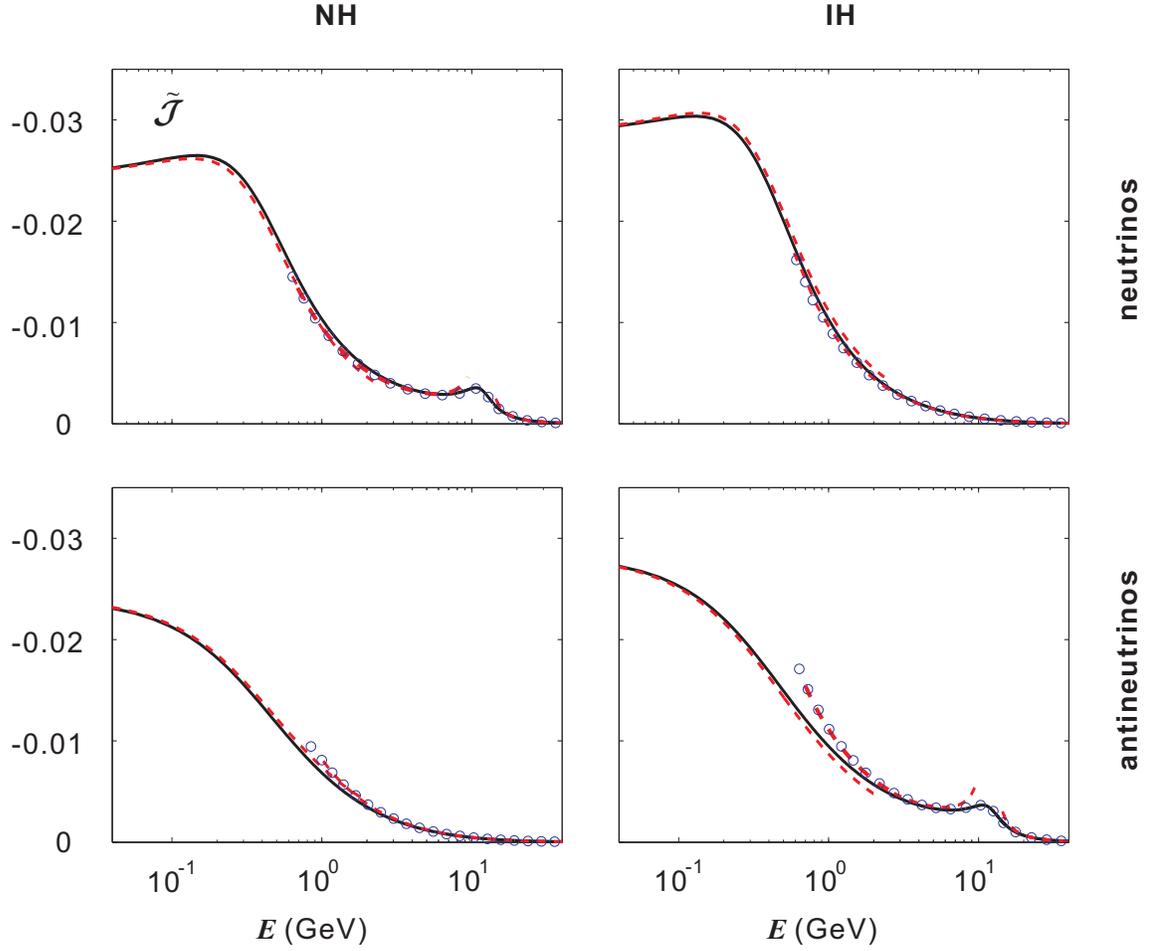}
\end{overpic}\caption{The analytical results of
Eqs. \eqref{Jarlskog2} and \eqref{Jarlskog3} (dashed line), Eq.
\eqref{Jarlskog1} (circle curve) and numerical results (solid line)
of the Jarlskog invariant. All the input parameters are taken the
same as in Fig. 2.}
\end{figure}

\begin{figure}[!h]
\center\label{mf}
\begin{overpic}[width=15cm]{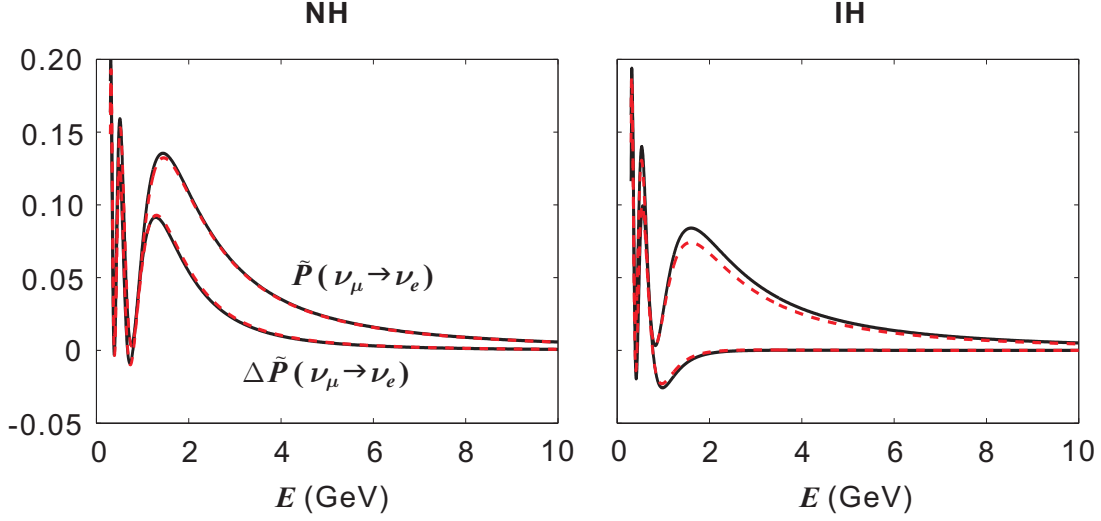}
\end{overpic}\caption{The analytical (dashed line)
and numerical (solid line) results of the oscillation probability
$\tilde{P}(\nu_\mu\rightarrow\nu_e)$ and the difference
$\Delta\tilde{P}(\nu_\mu\rightarrow\nu_e)$ with a length of baseline
$L=810$ km and matter density
$\rho^{}_\text{m}=2.8\;\text{g}/\text{cm}^3_{}$ in the NO$\nu$A
experiment \cite{NOvA}. The genuine flavor mixing parameters
($\theta^{}_1$, $\theta^{}_2$, $\theta^{}_3$ and
$\delta^{}_\text{KM}$) and mass-squired differences ($\Delta
m^2_{21}$ and $\Delta m^2_{31}$) are taken the same as in Fig. 2.}
\end{figure}

\begin{figure}[!h]
\center\label{mf}
\begin{overpic}[width=15cm]{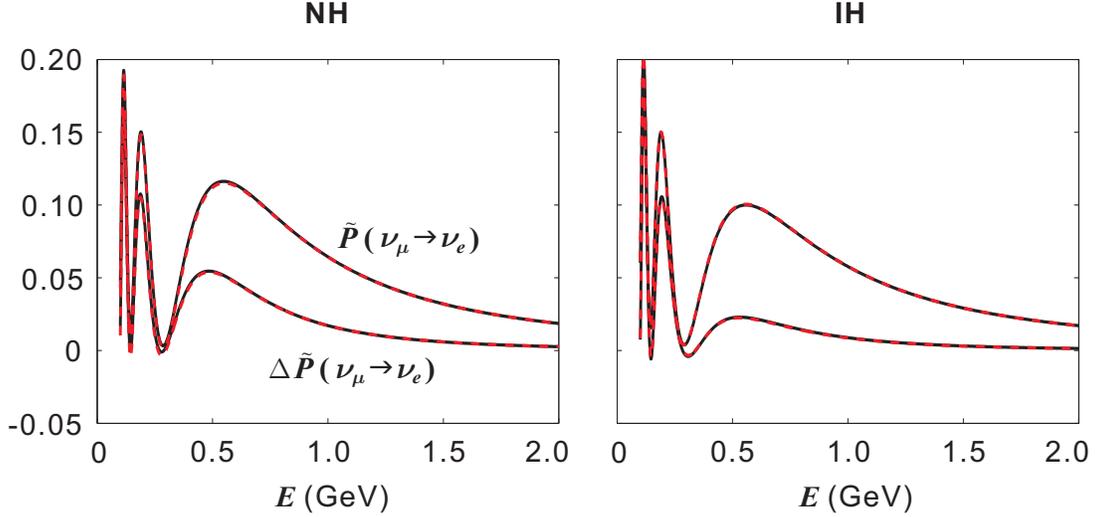}
\end{overpic}\caption{The analytical (dashed line) and
numerical (solid line) results of the oscillation probability
$\tilde{P}(\nu_\mu\rightarrow\nu_e)$ and the difference
$\Delta\tilde{P}(\nu_\mu\rightarrow\nu_e)$ in the Hyper-K experiment
\cite{HyperK}. Compared with Fig. 4, a length of baseline $L=295$ km
and matter density $\rho^{}_\text{m}=2.6\;\text{g}/\text{cm}^3_{}$
are taken. The genuine flavor mixing parameters and mass-squired
differences are taken the same as in Fig. 2.}
\end{figure}

\end{document}